\documentclass[conference]{IEEEtran}
\IEEEoverridecommandlockouts
\usepackage{cite}
\usepackage{amsmath,amssymb,amsfonts}
\usepackage{algorithmic}
\usepackage{graphicx}
\usepackage{textcomp}
\usepackage{xcolor}
\usepackage{graphicx}
\usepackage{array}
\usepackage{booktabs}
\usepackage{enumitem}
\usepackage[linesnumbered,ruled,vlined]{algorithm2e}
\usepackage{graphicx} 
\usepackage{newtxtext,newtxmath}
\def\BibTeX{{\rm B\kern-.05em{\sc i\kern-.025em b}\kern-.08em
    T\kern-.1667em\lower.7ex\hbox{E}\kern-.125emX}}

\begin{document}

\title{Time Series Based Network Intrusion Detection using MTF-Aided Transformer}

\author{\IEEEauthorblockN{$^1$Poorvi Joshi, $^2$Mohan Gurusamy}
\IEEEauthorblockA{ \textit{Department of Electrical and Computer Engineering}\\ \textit{National University of Singapore, Singapore}\\
 Email: $^1$e1144005@u.nus.edu, $^2$gmohan@nus.edu.sg}}



\maketitle

\begin{abstract}
This paper introduces a novel approach to time series classification using a Markov Transition Field (MTF)-aided Transformer model, specifically designed for Software-Defined Networks (SDNs). The proposed model integrates the temporal dependency modeling strengths of MTFs with the sophisticated pattern recognition capabilities of Transformer architectures. We evaluate the model's performance using the InSDN dataset, demonstrating that our model outperforms baseline classification models particularly in data-constrained environments commonly encountered in SDN applications. We also highlight the relationship between the MTF and Transformer components, which leads to better performance, even with limited data. Furthermore, our approach achieves competitive training and inference times, making it an efficient solution for real-world SDN applications. These findings establish the potential of MTF-aided Transformers to address the challenges of time series classification in SDNs, offering a promising path for reliable and scalable analysis in scenarios with sparse data.

\end{abstract}


\section{Introduction}
The widespread use of data-driven applications has led to an exponential increase in network complexity and a growing demand for faster, more reliable communication among network nodes which is a challenge for traditional networking solutions based on manual configuration, hierarchical and distributed control. SDN (Software Define Network) redefines it by centralizing network control and building flexibility and programmability into the network. With SDN architecture,  separated control and  data plane, enables dynamic adaptation of networks according to real-time needs, thereby providing agility and efficiency \cite{kreutz2014software}.
As a result, there has been broad SDN adoption in data centers, enterprise networks, and cloud environments, whereby traffic flows, scales, and automates. However, this programmability also introduces security challenges, such as the potential for control plane manipulation and an increased attack surface due to dynamic flow management \cite{rahouti2022sdn}. The dynamic nature of  SDN, combined with its high volume of network flows, has opened new paths for attackers to exploit vulnerabilities in both the control and data planes. 

Modern cyber threats have grown more advanced, with attacks designed to bypass defenses, disrupt critical services, and remain hidden within legitimate traffic patterns. Traditional security solutions, which rely on predefined rules and manual oversight, struggle to keep pace with SDN’s dynamic architecture. These static approaches often fall short when identifying sophisticated attacks that manipulate the control plane or exploit the continuous reconfiguration of network flows \cite{rahouti2022sdn}.

In the literature, flow-based intrusion detection has been explored using neural networks, particularly Multi-Layer Perceptrons (MLPs). Enhancements to the resilient backpropagation neural network (ERBP) have improved MLP training by adjusting weights based on local gradient signs, increasing learning efficiency \cite{naoum2012enhanced}. Deep Convolutional Neural Networks (DCNNs) have also been applied to intrusion detection due to their ability to recognize spatial patterns in structured data \cite{naseer2018enhanced}. Building on these approaches, recent advancements focus on transforming network flow data into structured formats to enhance classification accuracy. Decision trees are used to rank features based on their relevance to detecting specific attack types, selecting only the most critical attributes \cite{9862964}. These features are arranged into a 2D matrix representation, preserving essential relationships within the data. A common strategy converts these matrices into RGB images, enabling models like Convolutional Neural Networks (CNNs) \cite{al2021convolutional, toldinas2021novel} and Vision Transformers (ViTs) \cite{9862964} to classify them. However, direct feature-to-image conversion can lead to information loss, as numerical mappings may not fully retain the relationships necessary for accurate classification. Additionally, such methods often fail to capture temporal variations in network traffic, limiting their effectiveness against evolving threats. 

To overcome these limitations, time-series and packet-level analysis methods are increasingly being explored, offering better adaptability to real-time attack patterns in dynamic environments. 
\begin{figure*}[ht]
    \centering
    \includegraphics[width=1\textwidth]{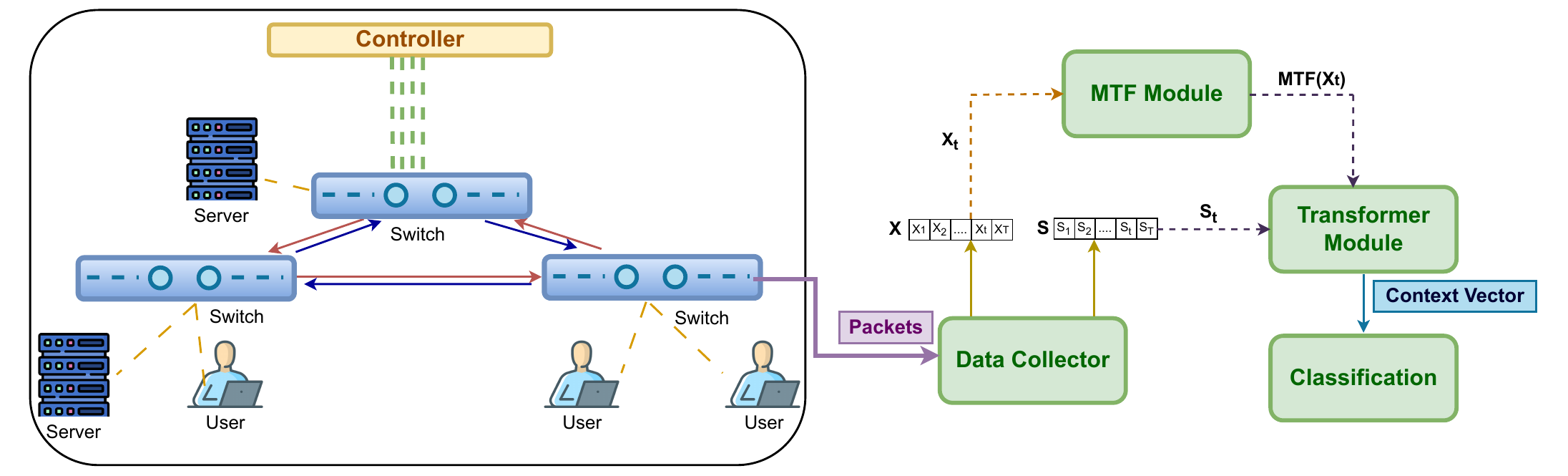}  
    \caption{System Architecture}
    \label{fig:system_architecture}
\end{figure*}

Time-series intrusion detection has been explored using deep learning models to capture temporal dependencies and identify deviations from normal patterns. Recurrent Neural Networks (RNNs) and their variants, such as Long Short-Term Memory (LSTM) and Gated Recurrent Unit (GRU), are commonly employed to model temporal contexts by predicting future values and detecting anomalies \cite{choi2021deep}. Convolutional Neural Networks (CNNs) have also been investigated for short-term time-series intrusion detection, leveraging hierarchical feature extraction \cite{wen2019time}. Temporal Convolutional Networks (TCNs) address some of the limitations of CNNs by ensuring causal convolutions \cite{bai2018empirical}. Hybrid models, such as ConvLSTM \cite{tayeh2022attention}, combine CNNs and LSTMs to simultaneously capture both spatial and temporal dependencies, while Transformer-based models are highly effective in capturing long-range dependencies through their attention mechanisms \cite{choi2021deep}.

However, each of the above approaches comes with its own set of challenges. RNN-based models are often computationally inefficient due to their sequential nature, which makes them less suitable for large-scale time-series data. CNNs typically require transforming sequential data into images, which can result in a loss of contextual integrity. While TCNs offer improvements over CNNs, they still struggle to preserve long-term dependencies effectively. Transformer models, though powerful, require a proper positional encoding when applied to time-series data, further it requires higher data set to learn the pattern, so there is need for proper time series data representation. These challenges become particularly problematic in SDN applications, where precise intrusion detection with minimal latency is essential for optimal performance.


We propose a fine-grained intrusion detection framework tailored for SDN by leveraging a Multivariate Markov Transition Field (MTF)-aided Transformer model to capture both temporal and spatial dependencies in network traffic. 
 Let $
\textbf{X} = \{X_1, X_2, ..., X_t, ..., X_T\}
$ denote the time-series data over \( T \) time slots, where each slot has a duration of \( \tau \). The network traffic at time slot \( t \) is represented as, \(
X_t = \{x_{t1}, x_{t2}, ..., x_{tl}, ..., x_{tL} \}
\), where \( x_{tl} \) corresponds to the time-series data of packet flow on the \( l^{th} \) link in the network.  Each time series corresponds to a specific link, characterized by a source, destination IP address, and protocol type. To model the network topology, we define a source-destination interaction matrix, \(
\textbf{S} = \{S_1, S_2, ..., S_t, ..., S_T\}
\), where \( S_t \in \mathbb{R}^{N \times N \times P} \) represents the network connectivity in time slot \( t \) between \( N \) source and destination IP addresses. Each entry in \( S_t \) captures both link existence and protocol-specific information, considering one-hot encoding for the protocols, where \(P\) is the number of distinct protocols. The spatial matrix remains fixed within each \( \tau \) duration, assuming no significant topology changes in short intervals.  


The performance evaluation is conducted using the InSDN \cite{elsayed2020insdn} dataset, where time-series data from SDN switches are transformed into Markov Transition Field (MTF) representations to capture transition dynamics. These structured representations are then processed using a Transformer-based model, which integrates spatial dependencies from \( S_t \) to enhance link-level intrusion detection. The combination of temporal evolution in packet flow (via MTF) and spatial dependencies in the network topology (via \( S_t \)) enhances the system’s ability to detect anomalies in real time. The evaluation considers key performance metrics—precision, recall, and F1-score—to assess the effectiveness of the proposed model. Even under 40\% data loss, the MTF-aided Transformer consistently outperforms baseline methods, achieving up to 26.5\% higher classification performance compared to traditional models. 

Contributions are as follow: 
\begin{itemize}
    \item We introduce an MTF-based encoding mechanism to transform raw time-series data into structured representations, enabling efficient intrusion detection through a Transformer encoder. 
    \item We incorporate a spatial matrix that captures source-destination and protocol-level interactions from SDN switches, allowing the model to analyze spatial-temporal relationships and detect anomalies at a fine-grained level.
    \item Our model achieves superior multi-class intrusion detection compared to baseline algorithms, under sparse data.
\end{itemize}

\section{System Architecture}

Our system architecture is built around three main components as shown in Fig.~\ref{fig:system_architecture}: data collection, MTF preprocessing, and stacked transformer-based classification. The first component, data collection, involves monitoring network traffic using switches that capture packet flows over a defined time span. This time span is further divided into smaller sub-slots of duration \( \tau \), during which we extract source and destination IP addresses to construct a space matrix over \( t \) time slots. We use the InSDN dataset \cite{elsayed2020insdn}, which includes both TCP and UDP traffic. To effectively represent different protocol flows, we apply one-hot encoding, where 00 indicates no flow, 01 represents a TCP flow, and 10 represents a UDP flow. This encoding helps us generate a space matrix \( S_t \) of size \( N \times N \times 2 \), where \( N \) is the number of network nodes.  

The second component, MTF (Markov Transition Field) preprocessing, transforms the collected multivariate time-series data \( X_t \). This time-series data captures how packet flows evolve over different time slots, providing a structured representation of network activity. The MTF module applies transformation techniques to each time-series of every network link in a given time slot, generating MTF(\( X_t \)) as an output. This transformation preserves important temporal relationships, making it easier to identify patterns associated with normal and malicious traffic. The third component, stacked transformer-based classification with embedding, is where the actual detection happens. Here, the processed MTF(\( X_t \)) and the space matrix \( S_t \) are fed into a deep-learning-based transformer model. A more detailed discussion of the Stacked Transformer Module, including its architectural components and the role of embeddings, will be provided later ~\ref{stack}. Once the classification is complete, the final decision is sent to the network administrator for further action. 


\subsection{MTF Pre-processing Module}

The Markov Transition Field (MTF) preprocessing module plays a crucial role in transforming multivariate time series data into a form that effectively captures temporal dependencies. Given a multivariate time series \( X_t \) at time \( t \), represented as:
\begin{equation} \label{1}
    X_t = \{x_{t,1}, x_{t,2}, \dots, x_{t,L}\},
\end{equation}

where each \( x_{t,l} \) corresponds to a univariate time series for feature \( l \), the module constructs a Markov Transition Field for each feature. 

The process begins by adapting the quantization of each univariate time series \( x_{t,l} \). Unlike traditional uniform or Gaussian-based quantization, which relies on fixed breakpoints derived from a normal distribution’s inverse cumulative distribution function (CDF) as,
\begin{equation}
q_i = \Phi^{-1}\left(\frac{i}{Q}\right), 
\end{equation}
adaptive quantization learns the bin boundaries during training. This approach allows the quantization process to adjust dynamically to the specific data distribution of each time series, thus improving the representation quality.

The quantization bin boundaries \( \theta = \{\theta_1, \theta_2, \dots, \theta_Q\} \) are treated as learnable parameters and are updated using backpropagation will be,
\begin{equation}\label{2}
\theta_i^{(t+1)} = \theta_i^{(t)} - \eta \frac{\partial \mathcal{L}}{\partial \theta_i}, 
\end{equation}
where \( \eta \) is the learning rate and \( \mathcal{L} \) is the loss function. By optimizing these parameters during training, the quantization strategy becomes more tailored to the task. Once the quantization process is complete, the Markov Transition Matrix \( W_l \) is constructed for each feature \( x_{t,l} \). 

\begin{algorithm}[htbp]
\label{MTF}
\caption{MTF Preprocessing Module}
\KwIn{Multivariate time series \( X_t = \{x_{t,1}, x_{t,2}, \dots, x_{t,L}\} \)}
\KwOut{\( MTF(X_t) = \{M_{t,1}, M_{t,2}, \dots, M_{t,L}\} \)}

\For{each feature \( l \in \{1, 2, \dots, L\} \)} {
    \textbf{Adaptive Quantization:} \\
    Initialize quantization bin boundaries \( \theta_l = \{\theta_1, \theta_2, \dots, \theta_Q\} \) randomly \\
    \For{each time step \( t \) in \( x_{t,l} \)} {
        Quantize the value \( x_{t,l} \) based on \( \theta_l \) (Eq. ~\ref{1})
    }
    
    \textbf{Markov Transition Matrix Construction:} \\
    Initialize \( W_l \) as a \( Q \times Q \) matrix of zeros \\
    \For{each pair of consecutive time steps \( k, k+1 \)} {
        Transition from quantized state \( q_j \) at time \( k \) to \( q_i \) at time \( k+1 \) \\
        Update transition probability \( P(q_i | q_j) \) in \( W_l \) (Eq.~\ref{3})
    }
    
    Normalize \( W_l \) by rows \\
    
    \textbf{Markov Transition Field Construction:} \\
    Initialize \( M_{t,l} \) as a matrix with elements \( P(q_{t_i} | q_{t_j}) = 0 \) \\
    \For{each pair of time steps \( t_j, t_i \)} {
        Transition from quantized state \( q_{t_j} \) at time \( t_j \) to \( q_{t_i} \) at time \( t_i \) \\
        Update transition probability \( P(q_{t_i} | q_{t_j}) \) in \( M_{t,l} \) (Eq. ~\ref{4})
    }
    
    Normalize \( M_{t,l} \) by rows \\
    
    \textbf{Dimensionality Reduction (Gaussian Blurring):} \\
    Apply Gaussian blurring to \( M_{t,l} \) to reduce dimensions (Eq. ~\ref{5}) \\
    
    Store \( M_{t,l} \) as part of \( MTF(X_t) \)
}

\KwRet{MTF matrices \( MTF(X_t) = \{M_{t,1}, M_{t,2}, \dots, M_{t,L}\} \)}
\end{algorithm}

This matrix captures the transition probabilities between quantized states, where \( P(q_i | q_j) \) represents the probability of transitioning from state \( q_j \) at time \( k \) to state \( q_i \) at time \( k+1 \). This probability is computed as,
\begin{equation}\label{3}
P(q_i | q_j) = \frac{\text{count}(x_k \in q_j, x_{k+1} \in q_i)}{\sum_{j=1}^Q \text{count}(x_k \in q_j)},
\end{equation}
The result is a Markov Transition Field \( M_{t,l} \) for each feature, which extends the transition matrix across time. For the multivariate time series \( X_t \), we obtain \( L \) MTF matrices will be given as,
\begin{equation}\label{4}
MTF(X_t) = \{M_{t,1}, M_{t,2}, \dots, M_{t,L}\}, 
\end{equation}
which collectively capture the temporal dependencies across all features. To improve computational efficiency and retain important transition information, dimensionality reduction is applied to each MTF matrix \( M_{t,l} \) using Gaussian blurring. The Gaussian kernel, will be defined as,

\begin{equation}\label{5}
G(x_1, x_2) = \frac{1}{2\pi\sigma^2} \exp\left(-\frac{x_1^2 + x_2^2}{2\sigma^2}\right), 
\end{equation}

smooths the MTF matrix, which reduces its size while preserving the critical transition patterns. This makes the MTF representation more manageable and suitable for further analysis, such as anomaly detection. The algorithm for the MTF preprocessing module is discussed in Algorithm ~\ref{MTF}, which outlines the step-by-step procedure for generating the MTF matrices.

\subsection{Stacked Transformer Algorithm} \label{stack}
\begin{figure*}[ht]
    \centering
    \includegraphics[width=0.8\textwidth]{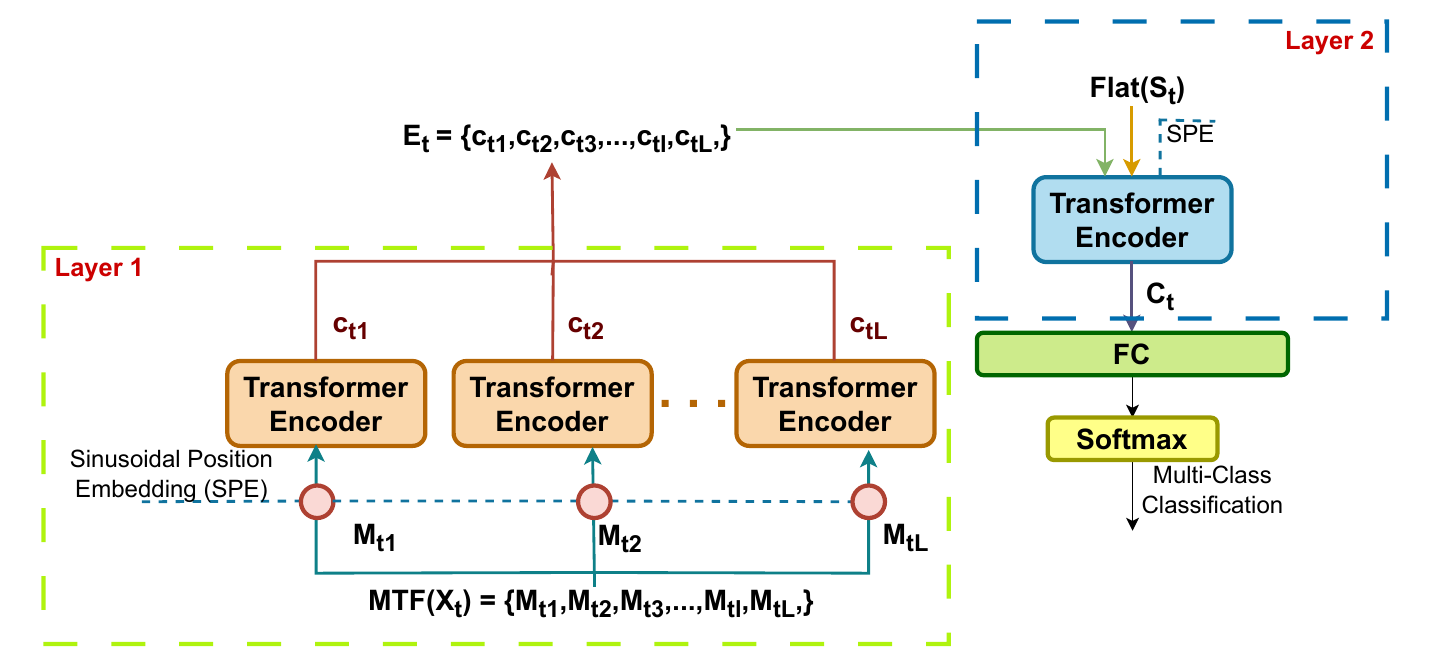} 
    \caption{Stacked Transformer Architecture}
    \label{trans}
\end{figure*}

The Stacked Transformer Module as shown Fig. ~\ref{trans} leverages two hierarchical Self-Attention Transformer Encoder layers to capture both temporal dependencies within individual features and structural relationships across features for multivariate time series classification.

\subsubsection{First Transformer Module: Feature-wise MTF Embedding}
Given a multivariate time series \( X_t = \{x_{t,1}, x_{t,2}, \dots, x_{t,L}\} \), where \( L \) represents the number of features, each Markov Transition Field (MTF) \( M_{t,l} \) (\( l \in \{1, 2, \dots, L\} \)) is processed independently. Each \( M_{t,l} \) is derived from the corresponding time series \( x_{t,l} \). Sinusoidal positional embeddings are added to the input MTF embeddings to provide temporal information:

\begin{equation} \label{eq:pos1}
PE(pos, 2i) = \sin\left( \frac{pos}{10000^{2i/d_{\text{model}}}} \right)
\end{equation}
\begin{equation} \label{eq:pos2}
PE(pos, 2i+1) = \cos\left( \frac{pos}{10000^{2i/d_{\text{model}}}} \right)
\end{equation}

where \( pos \) is the position index, \( i \) is the dimension index, and \( d_{\text{model}} \) is the embedding dimension. The input \( M_{t,l} \) is initially embedded into a vector of size \( (\tau/Q) \times 1 \), where \( \tau \) is the time series length, and \( Q \) is the number of quantization bins used to create the MTF.

\begin{algorithm}[htbp]
\caption{Stacked Transformer for Multivariate Time Series Classification}
\label{alg:stacked_transformer}
\KwIn{Multivariate time series \( MTF(X_t) = \{M_{t,1}, M_{t,2}, \dots, M_{t,L}\} \) and structural information \( S_t \)}
\KwOut{Predicted class probabilities \( Y_t \)}

\textbf{Initialize:} weight matrices \( W_Q, W_K, W_V, W_Q', W_K', W_V', W_o \) and bias vectors \( b_Q, b_K, b_V, b_Q', b_K', b_V', b_o \)

\textbf{First Transformer Module:}

\For{each feature \( l \in \{1, 2, \dots, L\} \)}{
    Compute \( Q_l = M_{t,l} W_Q + b_Q \) \quad \text{(Eq. \ref{eq:query})}\\
    Compute \( K_l = M_{t,l} W_K + b_K \) \quad \text{(Eq. \ref{eq:key})}\\
    Compute \( V_l = M_{t,l} W_V + b_V \) \quad \text{(Eq. \ref{eq:value})}\\
    Compute attention scores: \( A_l = \text{softmax}\left(\frac{Q_l K_l^T}{\sqrt{d_k}}\right) \) \quad \text{(Eq. \ref{eq:attention})}\\
    Compute context vector: \( C_{t,l} = A_l V_l \) \quad \text{(Eq. \ref{eq:context})}
}

Concatenate context vectors: \( E_t = \{C_{t,1}, C_{t,2}, \dots, C_{t,L}\} \) \quad \text{(Eq. \ref{eq:feature_representation})}\\

\textbf{Second Transformer Module:}

Flatten structural information \( S_t \) to \( L \times 1 \)\\
Concatenate \( E_t \) and \( S_t \): \( \hat{E}_t = \{E_t, S_t\} \) \quad \text{(Eq. \ref{eq:concat})}\\

\For{each feature in \( \hat{E}_t \)}{
    Compute \( Q_s = \hat{E}_t W_Q' + b_Q' \) \quad \text{(Eq. \ref{eq:query_s})}\\
    Compute \( K_s = \hat{E}_t W_K' + b_K' \) \quad \text{(Eq. \ref{eq:key_s})}\\
    Compute \( V_s = \hat{E}_t W_V' + b_V' \) \quad \text{(Eq. \ref{eq:value_s})}\\
    Compute attention scores: \( A_s = \text{softmax}\left(\frac{Q_s K_s^T}{\sqrt{d_k}}\right) \) \quad \text{(Eq. \ref{eq:attention_s})}\\
    Compute context vector: \( C_t = A_s V_s \) \quad \text{(Eq. \ref{eq:context_s})}
}

Compute output probabilities: \( Y_t = \text{softmax}(W_o C_t + b_o) \) \quad \text{(Eq. \ref{eq:output})}\\
Compute loss: \( \mathcal{L} = - \sum_{c=1}^{C} y_c \log(\hat{y}_c) \) \quad \text{(Eq. \ref{eq:loss})}\\
Update parameters using backpropagation

\end{algorithm}

For each \( M_{t,l} \), Query (\( Q_l \)), Key (\( K_l \)), and Value (\( V_l \)) matrices are computed as:

\begin{equation} \label{eq:query}
Q_l = M_{t,l} W_Q + b_Q
\end{equation}
\begin{equation} \label{eq:key}
K_l = M_{t,l} W_K + b_K
\end{equation}
\begin{equation} \label{eq:value}
V_l = M_{t,l} W_V + b_V
\end{equation}

where \( W_Q \), \( W_K \), \( W_V \) are weight matrices, and \( b_Q \), \( b_K \), \( b_V \) are bias vectors. Attention scores are calculated as:

\begin{equation} \label{eq:attention}
A_l = \text{softmax}\left(\frac{Q_l K_l^T}{\sqrt{d_k}}\right)
\end{equation}

where \( d_k \) is the dimension of the key vectors. The attention-weighted values produce the context vector \( C_{t,l} \):

\begin{equation} \label{eq:context}
C_{t,l} = A_l V_l
\end{equation}

The context vectors from all features are concatenated to form the feature representation \( E_t \):

\begin{equation} \label{eq:feature_representation}
E_t = \{C_{t,1}, C_{t,2}, \dots, C_{t,L}\}
\end{equation}

\subsubsection{Second Transformer Module: Combined Representation Learning}
The second transformer integrates \( E_t \) with structural information \( S_t \). \( S_t \) is an \( N \times N \times 2 \) tensor representing the structural relationships between the \( N \) time series (where \( L = N^2 \) after flattening). \( S_t \) is flattened to an \( L \times 1 \) vector. The input to the second transformer is the concatenation of \( E_t \) and \( S_t \):

\begin{equation} \label{eq:concat}
\hat{E}_t = \{E_t, S_t\}
\end{equation}

The same self-attention mechanism as in the first transformer is applied to \( \hat{E}_t \):

\begin{equation} \label{eq:query_s}
Q_s = \hat{E}_t W_Q' + b_Q'
\end{equation}
\begin{equation} \label{eq:key_s}
K_s = \hat{E}_t W_K' + b_K'
\end{equation}
\begin{equation} \label{eq:value_s}
V_s = \hat{E}_t W_V' + b_V'
\end{equation}
\begin{equation} \label{eq:attention_s}
A_s = \text{softmax}\left(\frac{Q_s K_s^T}{\sqrt{d_k}}\right)
\end{equation}
\begin{equation} \label{eq:context_s}
C_t = A_s V_s
\end{equation}

where \( W_Q' \), \( W_K' \), \( W_V' \) are weight matrices and \( b_Q' \), \( b_K' \), \( b_V' \) are bias vectors. \( C_t \) is the final contextualized representation.

The final context vector \( C_t \) is passed through a fully connected layer followed by a softmax function for multi-class classification:

\begin{equation} \label{eq:output}
Y_t = \text{softmax}(W_o C_t + b_o)
\end{equation}

where \( W_o \) is the output weight matrix and \( b_o \) is the output bias. \( Y_t \) represents the predicted probability distribution over the classes. The model is trained using the categorical cross-entropy loss function:

\begin{equation} \label{eq:loss}
\mathcal{L} = - \sum_{c=1}^{C} y_c \log(\hat{y}_c)
\end{equation}

where \( y_c \) is the true label (one-hot encoded), \( \hat{y}_c \) is the predicted probability for class \( c \), and \( C \) is the number of classes.

The hierarchical self-attention mechanism allows the model to learn complex relationships and improve classification performance on multivariate time series data. The use of MTFs provides a robust representation of the underlying dynamics of the time series, further enhancing the model's capabilities. This is mentioned in Algorithm 2.

\section{Results and Discussion}

We evaluate our proposed stacked transformer model using the SDN dataset "InSDN" \cite{elsayed2020insdn}, which includes various network attacks. The dataset consists of multiple categories of cyber threats, including Denial of Service (DoS) attacks, which were executed using LOIC, Slowhttptest, HULK, Torshammer, Nping, and the Metasploit framework. It also includes Distributed Denial of Service (DDoS) attacks performed using Hping3, as well as web attacks generated using Metasploit and SQLmap. Additionally, password brute-forcing attacks were carried out with Burp Suite, Hydra, and Metasploit. Probe attacks were conducted using Nmap and Metasploit, while exploitation-based attacks were executed using the Metasploit framework \cite{elsayed2020insdn}.

For the embedding process, we first extract flow features from the dataset and transform them into structured representations before feeding them into the stacked transformer model. To simulate data loss, we introduce different levels of data accessibility by randomly removing portions of the flow data, mimicking the effects of missing data in real-world scenarios. These levels of data accessibility are set to 100\%, 80\%, and 60\%, representing the percentage of available data during training and testing. We split the dataset into training and testing subsets using a standard 80-20 split, ensuring a fair evaluation of the model's performance. The experiments were conducted on a high-performance workstation equipped with an Intel Xeon W-3400 processor, 128 GB DDR5 RAM, and NVIDIA RTX A6000 GPUs. The system was operated on Windows 11 Pro for Workstations, ensuring a robust computational environment for training and testing our model. The hyperparameters used in our model are detailed in Table~\ref{HYPERPARAMETER}.

\begin{table}[ht]
\centering
\caption{Hyperparameter Settings for MTF-aided Transformer Model}
\label{HYPERPARAMETER}
\begin{tabular}{|l|c|}
\hline
\textbf{Hyperparameter} & \textbf{Value} \\
\hline
Learning Rate & 0.0005 \\
Optimizer & AdamW \\
Batch Size & 128 \\
Number of Attention Heads & 8 \\
Hidden Layer Size & 512 \\
Dropout Rate & 0.2 \\
Number of Transformer MLP Layers & 4 \\
Fully Connected Layers & 2 \\
Activation Function & ReLU \\
Weight Decay & 0.01 \\
Gradient Clipping & 1.0 \\
\hline
\end{tabular}
\end{table}

\begin{figure}[ht]
    \centering
    \includegraphics[width=0.45\textwidth]{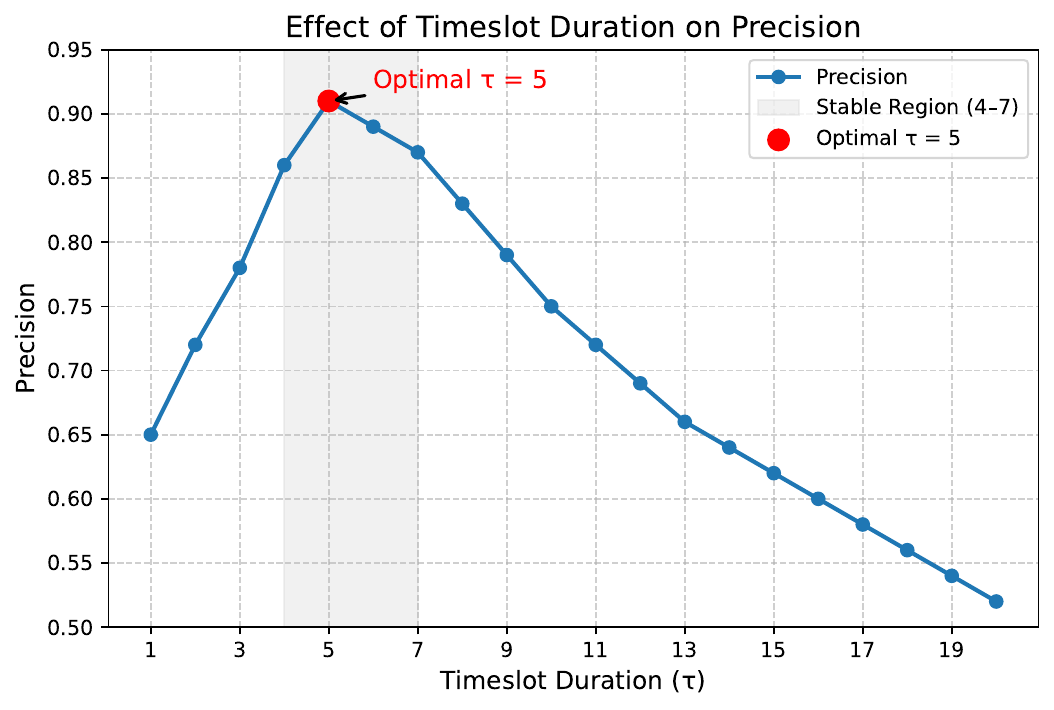} 
    \caption{Precision variation with respect to \(\tau\)}
    \label{fig:precision_tau}
\end{figure}

We evaluate our model's performance using key metrics such as F1 score, precision, recall, training time, and inference time. One important parameter in our model is \(\tau\), which represents the time slot duration and plays a crucial role in determining precision. Increasing \(\tau\) affects precision because our model assumes the network remains static within that period, while in reality, the network behavior is dynamic. Conversely, reducing \(\tau\) also lowers precision, though to a lesser extent, as the model may struggle to distinguish patterns due to the limited time series data available in shorter slots. Through extensive experimentation, we observed that setting \(\tau\) between 4 and 7 times the minimum flow duration provides stable performance across different runs. Based on these findings, we selected \(\tau = 5\) times the minimum flow duration for our final model. Fig. ~\ref{fig:precision_tau} presents how precision varies as \(\tau\) changes from 1 to 20 times the minimum flow duration.

We compare our model with traditional classification models such as k-Nearest Neighbors (KNN) \cite{zhang2017efficient}, Random Forest \cite{belgiu2016random}, LSTM \cite{said2020network}, and Donut \cite{xu2018unsupervised} for multi-class classification. These models aggregate time-series data for further processing. We evaluate them under different levels of data accessibility: 100\%, 80\%, and 60\%. Our model consistently outperforms others, especially when data accessibility is limited.

\begin{table}[ht]
\centering
\caption{Classification Performance Comparison}
\label{tab:multi_class}
\begin{tabular}{|l|c|c|c|}
\hline
\textbf{Model} & \textbf{Precision (\%)} & \textbf{F1 Score (\%)} & \textbf{Recall (\%)} \\
\hline
\multicolumn{4}{|c|}{\textbf{100\% Data Accessible}} \\
\hline
Proposed Model & \textbf{99.8} & \textbf{99.6} & \textbf{99.7} \\
KNN & 92.1 & 90.8 & 91.5 \\
Random Forest & 94.5 & 93.2 & 93.9 \\
LSTM & 95.2 & 94.0 & 94.6 \\
Donut & 96.3 & 95.0 & 95.7 \\
\hline
\multicolumn{4}{|c|}{\textbf{80\% Data Accessible}} \\
\hline
Proposed Model & \textbf{98.5} & \textbf{98.2} & \textbf{98.3} \\
KNN & 85.3 & 83.8 & 84.5 \\
Random Forest & 88.1 & 86.5 & 87.2 \\
LSTM & 90.4 & 88.9 & 89.6 \\
Donut & 92.2 & 90.8 & 91.5 \\
\hline
\multicolumn{4}{|c|}{\textbf{60\% Data Accessible}} \\
\hline
Proposed Model & \textbf{98.3} & \textbf{98.0} & \textbf{98.1} \\
KNN & 72.4 & 70.9 & 71.6 \\
Random Forest & 75.3 & 73.6 & 74.5 \\
LSTM & 80.1 & 78.6 & 79.3 \\
Donut & 85.0 & 83.5 & 84.3 \\
\hline
\end{tabular}
\end{table}

As shown in Table. ~\ref{tab:multi_class} The proposed method, MTF-aided Transformer, demonstrates superior performance, particularly in smaller datasets, compared to models like LSTM and Donut. This can be attributed to the unique capabilities of the Transformer architecture combined with the MTF. The Transformer model’s self-attention mechanism enables it to capture dependencies across the entire input sequence, allowing it to efficiently process and extract relevant features, even from sparse data. Unlike LSTM, which relies on sequential processing and can suffer from issues like vanishing gradients and overfitting when the data is limited, the Transformer’s non-sequential nature enables it to handle smaller datasets more robustly. This is especially important when training data is sparse, as the Transformer does not require large amounts of data to understand the context and relationships in the input.

When integrated with MTF, the Transformer model becomes even more effective in small-data environments. MTF allows the model to capture the probabilistic transitions between different states within the data, enriching the learned representations with context from prior states. This approach benefits small datasets by providing better generalization capabilities through its probabilistic structure. While LSTM focuses on long-term memory and Donut captures specific distributions in a non-sequential manner, the combination of MTF and Transformer is particularly powerful for small data because it can better handle the underlying temporal dependencies and complex patterns in the data, without being overwhelmed by noise or overfitting. The incorporation of MTF enhances the model’s ability to generalize, even in the presence of limited data, leading to improvements in precision, recall, and overall performance compared to LSTM and Donut, which do not leverage this type of transition-based contextualization.

\begin{table}[ht]
\centering
\caption{Training and Inference Time Comparison}
\label{tab:training_time}
\begin{tabular}{|l|c|c|}
\hline
\textbf{Model} & \textbf{Training Time (s)} & \textbf{Inference Time (ms)} \\
\hline
Proposed Model & 1200 & 8 \\
KNN & 900 & 15 \\
Random Forest & 1100 & 12 \\
LSTM & 2500 & 20 \\
Donut & 2800 & 18 \\
\hline
\end{tabular}
\end{table}

As shown in Table~\ref{tab:training_time}, the training time for the proposed MTF-aided Transformer model is 1200 seconds, which, while longer than simpler models like KNN (900s) and Random Forest (1100s), remains significantly shorter than more complex architectures such as LSTM (2500s) and Donut (2800s). The extended training time for LSTM is due to its sequential nature, which processes time steps recursively, while Donut’s complexity arises from its deeper network layers and probabilistic reconstruction mechanism. In contrast, our model leverages MTF to extract a structured time series representation by capturing transition probabilities within quantile bins, enabling efficient feature extraction and reducing the computational overhead.
Regarding inference time, the Transformer model performs efficiently with a response time of 8 milliseconds, which is quicker than both LSTM (20ms) and Donut (18ms). This is because the Transformer model leverages its self-attention mechanism, which can be parallelized, thus speeding up inference. In contrast, LSTM and Donut’s recurrent and specialized layers incur higher inference time due to the sequential processing of data and multi-step computations. Overall, while the Transformer requires slightly more training time than the simpler models, its inference time is faster, making it suitable for real-time applications. 

\begin{table}[ht]
\centering
\caption{Ablation Study Performance Comparison}
\label{tab:ablation_study}
\begin{tabular}{|l|c|c|c|}
\hline
\textbf{Model Setup} & \textbf{Precision(\%)} & \textbf{F1 Score(\%)} & \textbf{Recall(\%)} \\
\hline
MTF - Transformer & \textbf{99.8} & \textbf{99.6} & \textbf{99.7} \\
Without MTF & 95.3 & 94.8 & 94.9 \\
Without Transformer & 93.6 & 92.5 & 93.2 \\
\hline
\end{tabular}
\end{table}

In the ablation study, we analyzed the performance of the MTF-aided Transformer by comparing it with its variants. The full model, which incorporates both the MTF component and the Transformer architecture, outperformed the others, achieving precision, F1 score, and recall values of 99.8\%, 99.6\%, and 99.7\%, respectively. When we removed the MTF component and relied solely on the Transformer, performance dropped significantly, with precision, F1 score, and recall of 95.3\%, 94.8\%, and 94.9\%. Similarly, replacing the Transformer with a simpler feedforward network while keeping the MTF component also resulted in a noticeable decline in performance, with precision, F1 score, and recall of 93.6\%, 92.5\%, and 93.2\%. These results clearly demonstrate that the combined use of MTF and Transformer is critical for maximizing performance. The MTF component helps capture temporal dependencies, while the Transformer excels at handling complex patterns in the data. Removing either one of these components leads to a noticeable drop in accuracy, reinforcing the importance of their integration for high-quality predictions.

\section{Conclusion}
In this paper we developed a novel approach for time series classification through the MTF-aided Transformer model, which demonstrates substantial performance improvements over existing methods such as KNN, Random Forest, LSTM, and Donut, particularly when data is limited. This is achieved by leveraging Markov Transition Fields (MTFs) to capture temporal dependencies and combining them with the Transformer architecture's ability to model complex patterns. The model was tested using publicly available SDN dataset InSDN. Our approach achieves a performance improvement of up to 26.5\% better precision compared to the baseline techniques, with up to 40\% data loss. Additionally, the model outperforms the baseline models by achieving 50.76\% faster detection on average. Further, ablation study shows the complementary nature of MTFs and Transformers, highlighting the importance of each component. The marked performance drop when either is removed underscores their essential roles in achieving optimal classification results. Additionally, our model maintains competitive training and inference times, making it a practical solution for real-world applications. These results reinforce the potential of MTF-aided Transformers in addressing the challenges of time series classification, particularly in environments with sparse data.

\section*{Acknowledgment}
This research is supported by A*STAR, CISCO Systems
(USA) Pte. Ltd., and the National University of Singapore
under its Cisco-NUS Accelerated Digital Economy Corporate
Laboratory (Award I21001E0002).

\bibliographystyle{IEEEtran}
\bibliography{ref}

\end{document}